\documentclass{jps-cp}
\usepackage{txfonts} 

\title{Doubly Heavy Baryons Expanded in $\boldmath{1/m_Q}$}

\author{Takayuki \textsc{Matsuki}$^{1,2}$}

\inst{$^{1}$Tokyo Kasei University, 1-18-1 Kaga, Itabashi, Tokyo 173-8602, Japan \\
$^{2}$Theoretical Research Division, Nishina Center, RIKEN, Wako, Saitama 351-0198, Japan
}

\email{matsuki@tokyo-kasei.ac.jp}

\recdate{\today}

\abst{Starting from the semirelativistic Hamiltonian for a doubly heavy baryon system ($QQq$) with Coulomb and linear confining scalar potentials, and operating the naive Foldy-Wouthuysen-Tani transformation on two heavy quarks, I construct a formulation how to calculate mass spectra and wave functions of doubly heavy baryons. Based on our formulation, their masses and wave functions are expanded in $1/m_Q$ with a heavy quark mass $m_Q$ and the $\lambda$ mode lowest-order equation is examined carefully to obtain a complete set of eigenfunctions.
I claim that the $\rho$ mode wave function should be given by other nonrelativistic methods, e.g., the Godfrey-Isgur model to give a complete wave function for a doubly heavy baryon.
}

\kword{heavy baryon, heavy quark symmetry, spectrum}

\begin{document}
\maketitle

\section{Introduction}
Heavy quark symmetry can be respected in hadrons in which at least one heavy quark is included and other quarks are light quarks. The simplest case is a heavy-light meson and the next will be a doubly heavy baryon $QQq$ because compared with $Qqq$ baryon, there is only one light quark involved. Treating two heavy quark is rather easier compared with two light quarks. The heavy-light meson has been extensively studied by us in Ref. \cite{Matsuki:1997da} and also by many groups (see Ref. \cite{Swanson:2006st} for a review).

In this article, we will describe how to treat doubly heavy baryons using the method adopted in the former paper \cite{Matsuki:1997da} where heavy-light mesons have been studied.
Starting from the semirelativistic Hamiltonian for a doubly heavy baryon system ($QQq$) with Coulomb and linear confining scalar potentials, and operating the naive Foldy-Wouthuysen-Tani transformation on the heavy quarks, I expand all the physical quantities, Hamiltonian $H$, wave function $\psi$, and energy eigenvalue $E$, in $1/m_Q$ with heavy quark mass $m_Q$.
However, as for the so-called $\rho$ mode wave function, we need to use other methods to calculate. Godfrey-Isgur relativised potential model is one choice. This is because by using the FWT transformation, we cannot obtain matrix structure for the $\rho$ mode Hamiltonian so that we cannot distinguish a couple of spin states of a heavy diquark.

\section{Formulation}
Using heaviness of $c$ or $b$ quarks compared with light quarks, $u, d,$ and $s$, we apply the method developed in Ref. \cite{Matsuki:1997da} to a doubly heavy baryon. In Ref. \cite{Matsuki:1997da}, we have used the Foldy-Wouthuysen-Tani (FWT) transformation to expand the system in $1/m_Q$. Since there are two heavy quarks in a doubly heavy baryon, we need to operate two kinds of the FWT transformation on $QQq$.

%
\begin{figure}[tbh]
\begin{center}
\scalebox{0.3}{\includegraphics{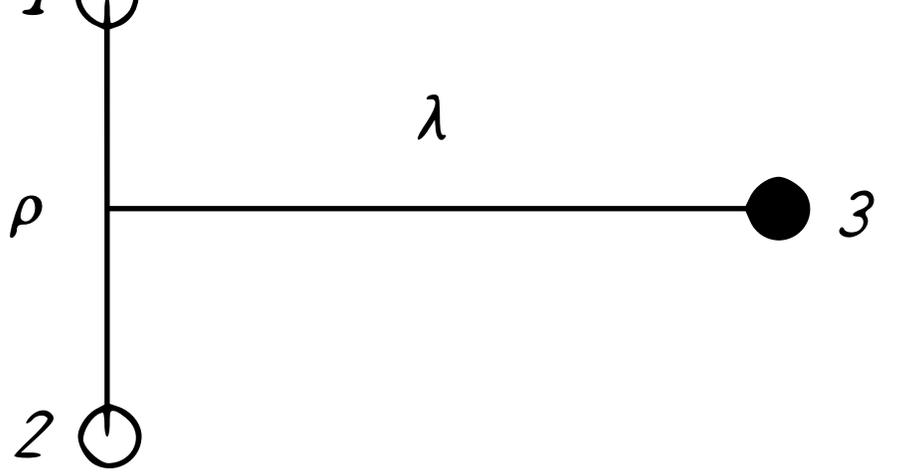}}
\caption{Definition of $\rho$ and $\lambda$. \label{fig1}}
\end{center}
\end{figure}
Instead of using the Cartesian coordinates $\vec r_1, \vec r_2$, and $\vec r_3$, we use the Jacobi relative coordinates $\vec \lambda$ and $\rho$.
If we regard quarks 1 and 2 are heavy with mass $m_1=m_2=m_Q$ and quark 3 is light with $m_3=m_q$, the Jacobi coordinate system can be simplified as,
\begin{eqnarray}
  \vec\rho = \frac{1}{\sqrt{2}}\left(\vec r_1 -\vec r_2\right),\quad
  \vec\lambda = \frac{1}{\sqrt{6}}\left(\vec r_1+\vec r_2-2\vec r_3\right). \label{eq4}
\end{eqnarray}

After expansion, we obtain the lowest order eigenvalue equation for the $\lambda$ mode wave function as,
\begin{eqnarray}
  \tilde H_0 = -\vec p_{\lambda'}\cdot\vec\alpha_3 + m_q\beta_3+2V(\lambda') +
  2\beta_3 S(\lambda'), \quad 
   \tilde H_0 \psi_0 = E_0\psi_0, \label{effH0}
\end{eqnarray}
where $\lambda^\prime=\lambda/\sqrt{6}$, one-gluon exchange potential $V(r)=-2\alpha_s/(3r)$, and confining linear potential $S(r)=r/a^2+b$. This equation expresses nothing but the one for interaction between a light quark and doubly heavy diquark, which is automatically derived from our formulation.

An angular part of a solution to Eq. (\ref{effH0}) is explicitly given as follows.
We define the following eigenfaunctions for Eq. (\ref{effH0}),
\begin{eqnarray}
  y_{jm}^k(\Omega) &=& \frac{1}{\sqrt{2(j+1)}}
  \left( {\begin{array}{*{20}{c}}{\sqrt {j + 1 - m} \;Y_{j + 1/2}^{m - 1/2}}\\
  { - \sqrt {j + 1 + m} \;Y_{j + 1/2}^{m + 1/2}}
  \end{array}} \right), \\
  y_{jm}^{-k}(\Omega) &=& \frac{1}{\sqrt{2j}}
  \left( {\begin{array}{*{20}{c}}{\sqrt {j + 1 - m} \;Y_{j + 1/2}^{m - 1/2}}\\
  { - \sqrt {j + 1 + m} \;Y_{j + 1/2}^{m + 1/2}}
  \end{array}} \right) = \left( {\vec \sigma\cdot \vec n} \right)\;y_{jm}^k(\Omega),
\end{eqnarray}
where $Y_j^m$ are spherical hamonics, and $k=j+1/2$. Here is a relation between $k$, $l$, and $j$ as
\begin{eqnarray}
  {\rm when~}l=k\pm \frac{1}{2}, \quad{\rm then~}k= \pm\left(j+\frac{1}{2}\right),
\end{eqnarray}
and $k$ is an eigenvalue of an operator $K = -\beta_q\left(\vec\Sigma_q\cdot \vec L_\lambda +1\right)$.
Then a general solution to Eq. (\ref{effH0}) is given by
\begin{eqnarray}
  \psi_{jm}^k &=& \frac{1}{r}
  \left( {\begin{array}{*{20}{c}}{u_k(r)}\\
  { - iv_k(r)\left( {\vec \sigma\cdot \vec n} \right)}
  \end{array}} \right) y_{jm}^k(\Omega ).
\end{eqnarray}
where functions $u_k(r)$ and $v_k(r)$ satisfy the following eigenvalue equation:
\begin{eqnarray}
  \left( {\begin{array}{*{20}{c}}{{m_q} + 2S + 2V}&{ - {\partial _r} + \frac{k}{r}}\\
  {{\partial _r} + \frac{k}{r}}&{ - {m_q} - 2S + 2V}
  \end{array}} \right)\left( {\begin{array}{*{20}{c}}
  {{u_k}(r)}\\
  {{v_k}(r)}
  \end{array}} \right) = E_0^k\left( {\begin{array}{*{20}{c}}
  {{u_k}(r)}\\
  {{v_k}(r)}
\end{array}} \right).,\label{eq:lambdaeigen}
\end{eqnarray}
which can be solved like in Ref. \cite{Matsuki:1997da} using the variational method.

\section{Physics related to heavy-light systems}
\begin{flushleft}
1) Relation among $L_\rho$, $s_\rho$, and state symbols
\end{flushleft}

Because the total wave function of a diquark should be antisymmetric for two heavy quarks, there is a relation between $L_\rho$ and $s_\rho$. First, I should mention that a diquark with the same two heavy quarks, $c$ or $b$, is flavor symmetric and color antisymmetric.
As for a combination of two heavy quarks, if a diquark has spin $s_\rho=0$, i.e., two heavy quarks have opposite spin directions, a spin wave function is antisymmetric. On the other hand, if a diquark has spin $s_\rho=1$, a spin wave function is symmetric.
When we denote spin as $s_\rho$, angular momentum as $L_\rho$, and parity as $P_\rho$ for a $\rho$ mode diquark, we have the following combinations,
\begin{eqnarray}
  &&\left(s_\rho = 0, \quad L_\rho = 1,3,\cdots, \quad P_\rho=- \right),
  \\
  &&\left(s_\rho = 1, \quad L_\rho = 0,2,\cdots, \quad P_\rho=+ \right). 
\end{eqnarray}
As you can see from these combinations, when the value of $L_\rho$ is given, the value of $s_\rho$ becomes unique.
Together with these quantum numbers, we need to consider quantum numbers coming from the $\lambda$ mode, i.e., principal quantum number $n_\lambda$, angular momentum $l_\lambda$, and parity $P_\lambda=(-)^{l_\lambda}$.

What kind of symbol should be adopted is an interesting problem. We list the three kinds of them.
\begin{eqnarray}
  {\rm  Ref.}[2] &:& \quad \left(N_\rho L_\rho n_\lambda l_\lambda\right)J^P, \\
  {\rm  Ref.}[3] &:& \quad \left(N_\rho L_\rho n_\lambda l_\lambda\right)_{j_\rho}J^P, \\
  {\rm Ours}&:& \quad \left(N_\rho L_\rho n_\lambda l_\lambda {k}\right)J^P.
\end{eqnarray}
where $P=(-)^{L_\rho+L_\lambda}$, $k=\pm(j_\rho+1)$, and the $\rho$ mode principal quantum number $N_\rho$.
Ref. \cite{Lu:2017meb} and ours are equivalent but Ref. \cite{Ebert:2002ig} lacks information on $\rho$ mode quantum numbers.

\begin{flushleft}
2) Threshold behaviors
\end{flushleft}

Let us consider the similarity of doubly heavy baryons to heavy-light mesons, especially to $D_s(0^+, 1^+)$, which have very narrow widths. This is described as follows: Assume that $SU(3)$ light meson and quark interaction is given by
\begin{eqnarray}
  \mathcal{L}_{\rm int} &=& \frac{g}{\sqrt{2}f_\pi}\bar q_i\gamma_\mu\gamma^5\partial^\mu\phi_{ij}q_j, \\
  (\phi)_{ij} &=& \sqrt 2 \left( {\begin{array}{*{20}{c}}
{\frac{1}{{\sqrt 2 }}{\phi _3} + \frac{1}{{\sqrt 6 }}{\phi _8}}&{{\pi ^ + }}&{{K^ + }}\\
{{\pi ^ - }}&{ - \frac{1}{{\sqrt 2 }}{\phi _3} + \frac{1}{{\sqrt 6 }}{\phi _8}}&{{K^0}}\\
{{K^ - }}&{{{\bar K}^0}}&{ - \frac{2}{{\sqrt 6 }}{\phi _8}}
\end{array}} \right), \\
  \left( {\begin{array}{*{20}{c}}
{{\pi ^0}}\\
\eta 
\end{array}} \right) &=& \frac{1}{{\sqrt {1 + {\epsilon ^2}} }}\left( {\begin{array}{*{20}{c}}
1&\epsilon \\
{ - \epsilon }&1
\end{array}} \right)\left( {\begin{array}{*{20}{c}}
{{\phi _3}}\\
{{\phi _8}}
\end{array}} \right)\quad {\rm or}\quad
  \left( {\begin{array}{*{20}{c}}
{{\phi _3}}\\
{{\phi _8}}
\end{array}} \right)
   = \frac{1}{{\sqrt {1 + {\epsilon ^2}} }}\left( {\begin{array}{*{20}{c}}
1& -\epsilon \\
{ \epsilon }&1
\end{array}} \right) \left( {\begin{array}{*{20}{c}}
{{\pi ^0}}\\
\eta 
\end{array}} \right).
\end{eqnarray}
Then, since $s$ quark ($i, j=3$) inside of $D_s$ couples to $\phi_8$ which is mixed with pion ($\pi^0$), heavy-light meson can decay into another heavy-light meson + $\pi^0$ with a small mixing parameter $\epsilon=1.0\times 10^{-2}$.

If $M(D_s(0^+) > M(D(0^-)) + M(K)$, then we would expect a broard decay width of the $D_s(0^+)$. However, what we have found is the oppsite situation so that the decay width of this state becomes very narrow because the allowed decay channel is $D_s(0^+)\to D_s(0^-)+\pi$ which occurs through very small $\pi^0-\eta$ coupling.

A similar process to this in doubly heavy baryons is given by, if $M(\Omega_{cc}({3}/{2}^+)) < M(\Xi_{cc}({1}/{2}^+)) + M(K)$,
\begin{eqnarray}
  \Omega_{cc}\left({3}/{2}^+\right)^+(ccs) \to \Omega_{cc}\left({1}/{2}^-\right)^+(ccs)+\pi^0.
\end{eqnarray}
if $M(\Omega_{cc}({3}/{2}^+)) > M(\Xi_{cc}({1}/{2}^+)) + M(K)$, we expect a broad decay width due to the existence of the process,
\begin{eqnarray}
  \Omega_{cc}\left({3}/{2}^+\right)^+(ccs) \to \Xi_{cc}\left({1}/{2}^-\right)^{++}(ccu)+K^{-}.
\end{eqnarray}
where $K$ has mass of 494-498 MeV, the quark content of $\Xi_{cc}\left({1}/{2}^+\right)^+$ is $ccd$, while that of $\Xi_{cc}\left({1}/{2}^+\right)^{++}$ is $ccu$.
Doubly heavy baryons have excitations in $\rho$ mode, i.e., the first excitation is given by $L_\rho L_\lambda =P_\rho S_\lambda$ with a total angular momentum $J^P=3/2^+$ and the ground states are given by $S_\rho S_\lambda$ with $J^P=1/2^-$.

\begin{flushleft}
2) Mixing angles in heavy-quark symmetry
\end{flushleft}

In principle,there could be mixing between staes with the same quantum number $J^P$. For instance, there may be mixing between states with $J^P=1/2^-$, e.g., ${(1S1p)1/{2^ - }(j_\lambda=1/2)}$, ${(1S1p)1/{2^ - }(j_\lambda=3/2)}$,  and ${(1P1s)1/{2^ - }(j_\lambda=1/2)}$.
We have checked whether there is mixing, e.g., between ${(1S1p)1/{2^ - }(j_\lambda=1/2)}$ and ${(1P1s)1/{2^ - }(j_\lambda=1/2)}$. It turns out that as long as we consider matrix elements with interactions of spin-spin and $LS$ coupling, we obtain no mixing between them, i.e., there is mixing between the states with the same $L_\lambda$ and different $L_\rho$ but no mixing between $S_\rho P_\lambda$ and $P_\rho S_\lambda$.
Because we use the heavy quark symmetry, we obtain mixing angles between certain states \`a la heavy-light mesons \cite{Matsuki:2010zy}. One example is given by,
\begin{eqnarray}
\left( {\begin{array}{*{20}{c}}
{(1S1p)1/{2^ - }}\\
{(1S1p)1/{2^{{\rm{'}} - }}}
\end{array}} \right) = \left( {\begin{array}{*{20}{c}}
{\cos \theta }&{ - \sin \theta }\\
{\sin \theta }&{\cos \theta }
\end{array}} \right)\left( {\begin{array}{*{20}{c}}
{(1S1p)1/{2^ - }(j_\lambda=1/2)}\\
{(1S1p)1/{2^ - }(j_\lambda=3/2)}
\end{array}} \right).
\end{eqnarray}
with $\theta=\arctan(1/\sqrt{2})=35.3^\circ$ in the heavy quark limit.

\section{Summary}

In this article, I have described how one can obtain the $\lambda$ mode wave function with an explicit angular part. Other physical quantities, good quantum number $K$, state symbols, possible threshold behaviors, and mixing angles among states with the same quantum numbers, have been discussed.

\section*{Acknowledgement}
I would like to thank Qi-Fang L${\ddot {\rm u}}$ of Huan Normal University for helpful discussions.

\end{document}